\documentclass[11pt,a4paper]{article}
\usepackage{jcappub}
\usepackage{graphics}
\usepackage{dcolumn}
\usepackage{amsfonts}
\usepackage{amssymb}
\usepackage{bm}
\usepackage{url}
\usepackage{subfigure}
\usepackage[sort&compress]{natbib}
\usepackage{txfonts}
\usepackage{color}

\begin{document}
\newcommand{\changeR}[1]{{#1}}

\newcommand{\TBB}{{{T_{\rm BB}}}}
\newcommand{\TCMB}{{{T_{\rm CMB}}}}
\newcommand{\Te}{{{T_{\rm e}}}}
\newcommand{\Teq}{{{T^{\rm eq}_{\rm e}}}}
\newcommand{\Ti}{{{T_{\rm i}}}}
\newcommand{\nB}{{{n_{\rm B}}}}
\newcommand{\nHe}{{{n_{\rm He}}}}
\newcommand{\nH}{{{n_{\rm H}}}}
\newcommand{\nHet}{{{n_{\rm ^3He}}}}
\newcommand{\nHt}{{{n_{\rm { }^3H}}}}
\newcommand{\nHtw}{{{n_{\rm { }^2H}}}}
\newcommand{\nBes}{{{n_{\rm { }^7Be}}}}
\newcommand{\nLis}{{{n_{\rm { }^7Li}}}}
\newcommand{\nLisi}{{{n_{\rm { }^6Li}}}}
\newcommand{\nS}{{{n_{\rm s}}}}
\newcommand{\Teff}{{{T_{\rm eff}}}}

\newcommand{\id}{{{\rm d}}}
\newcommand{\aR}{{{a_{\rm R}}}}
\newcommand{\bR}{{{b_{\rm R}}}}
\newcommand{\neb}{{{n_{\rm eb}}}}
\newcommand{\neql}{{{n_{\rm eq}}}}
\newcommand{\kB}{{{k_{\rm B}}}}
\newcommand{\EB}{{{E_{\rm B}}}}
\newcommand{\zmin}{{{z_{\rm min}}}}
\newcommand{\zmax}{{{z_{\rm max}}}}
\newcommand{\YBEC}{{{Y_{\rm BEC}}}}
\newcommand{\YSZ}{{{Y_{\rm SZ}}}}
\newcommand{\rhob}{{{\rho_{\rm b}}}}
\newcommand{\Ne}{{{n_{\rm e}}}}
\newcommand{\sigT}{{{\sigma_{\rm T}}}}
\newcommand{\me}{{{m_{\rm e}}}}
\newcommand{\nBB}{{{n_{\rm BB}}}}
\newcommand{\run}{{{\id \nS/\id\ln k}}}

\newcommand{\kD}{{{{k_{\rm D}}}}}
\newcommand{\KC}{{{{K_{\rm C}}}}}
\newcommand{\KdC}{{{{K_{\rm dC}}}}}
\newcommand{\Kbr}{{{{K_{\rm br}}}}}
\newcommand{\zdC}{{{{z_{\rm dC}}}}}
\newcommand{\zbr}{{{{z_{\rm br}}}}}
\newcommand{\aC}{{{{a_{\rm C}}}}}
\newcommand{\adC}{{{{a_{\rm dC}}}}}
\newcommand{\abr}{{{{a_{\rm br}}}}}
\newcommand{\gdC}{{{{g_{\rm dC}}}}}
\newcommand{\gbr}{{{{g_{\rm br}}}}}
\newcommand{\gff}{{{{g_{\rm ff}}}}}
\newcommand{\xe}{{{{x_{\rm e}}}}}
\newcommand{\alphafs}{{{{\alpha_{\rm fs}}}}}
\newcommand{\YHe}{{{{Y_{\rm He}}}}}
\newcommand{\SE}{{{{{dQ}}}}}
\newcommand{\SQ}{{{{{dz}}}}}
\newcommand{\SN}{{\dot{\mathcal{N}}}}
\newcommand{\Sn}{{{\mathcal{N}}}}
\newcommand{\muc}{{{{\mu_{\rm c}}}}}
\newcommand{\xc}{{{{x_{\rm c}}}}}
\newcommand{\xH}{{{{x_{\rm H}}}}}
\newcommand{\mT}{{{{\mathcal{T}}}}}
\newcommand{\Ob}{{{{\Omega_{\rm b}}}}}
\newcommand{\Or}{{{{\Omega_{\rm r}}}}}
\newcommand{\Ocdm}{{{{\Omega_{\rm cdm}}}}}
\newcommand{\mdm}{{{{m_{\rm WIMP}}}}}
\newcommand{\yg}{{{{y_{\gamma}}}}}
\newcommand{\mpci}{{{{{\rm Mpc}^{-1}}}}}

\title{Forecasts for CMB $\mu$ and $i$-type spectral distortion  constraints on the primordial power spectrum on scales
$8\lesssim k \lesssim 10^4 ~\rm{Mpc}^{-1}$  with the future Pixie-like experiments}

\author[a]{Rishi Khatri,}
\author[a,b]{Rashid A. Sunyaev}

\affiliation[a]{ Max Planck Institut f\"{u}r Astrophysik\\, Karl-Schwarzschild-Str. 1
  85741, Garching, Germany }
\affiliation[b]{Space Research Institute, Russian Academy of Sciences, Profsoyuznaya
 84/32, 117997 Moscow, Russia}
\date{\today}
\emailAdd{khatri@mpa-garching.mpg.de}
\abstract
{
Silk damping  at redshifts $1.5\times 10^4 \lesssim z \lesssim 2\times 10^6$
erases CMB anisotropies on scales  corresponding to the comoving
wavenumbers $8\lesssim k \lesssim 10^4 ~\rm{Mpc}^{-1}$ ($10^5 \lesssim \ell
\lesssim 10^8$). This dissipated
energy is gained by the CMB monopole, creating distortions from a blackbody
in the CMB spectrum of  the $\mu$-type and the $i$-type. We
study, using Fisher matrices, the constraints we can get from measurements
of these spectral distortions on the primordial power
spectrum from future experiments such as Pixie, and  how these
constraints change as we change the frequency resolution and the sensitivity of the experiment.  We show that the additional
information in the shape of the $i$-type distortions, in combination with
the $\mu$-type distortions,  allows us to break the
degeneracy between the amplitude and the spectral index of the power
spectrum on these scales {and leads to much  tighter constraints.} We quantify the information contained in
both the $\mu$-type distortions and the $i$-type distortions taking into
account the partial degeneracy with the $y$-type distortions and the
temperature of the blackbody part of the CMB. We also calculate the
constraints possible on the primordial power spectrum when the spectral
distortion information is combined with the CMB anisotropies measured by
the WMAP, SPT, ACT and Planck experiments.
}

\maketitle
\flushbottom

\section{The spectrum of CMB}
In the early Universe, at redshifts $z\gg 2\times 10^6$, there is
almost perfect thermal equilibrium between photons and electrons/baryons
which  maintains the spectrum of the cosmic microwave background (CMB)  to
be a blackbody spectrum even in the presence of enormous  energy injection
such as during electron-positron annihilation. This prediction of the standard big bang
cosmological model 
was confirmed by the Far Infrared
Absolute Spectrophotometer (FIRAS) instrument on the Cosmic Background
Explorer satellite (COBE) \cite{cobe} which found that the CMB is indeed
blackbody to high precision. If there is any energy injection  into (or
cooling of) the primordial plasma at
 $z\gtrsim 2\times 10^6$, Compton scattering is able to very quickly
 redistribute this excess (or deficit) of energy over the entire spectrum of photons restoring the
 equilibrium Bose-Einstein
 distribution \cite{sz1970} with chemical potential parameter $\mu$ and occupation number
 $n(x)=1/(e^{x+\mu}-1)$, where $x=h\nu/\kB T$, $h$ is Planck's
 constant, $\kB$ is Boltzmann's constant, $\nu$ is the photon frequency and
 $T$ is the temperature of photons and baryons. The chemical potential is
 in turn driven to zero by photon production \cite{sz1970} by
 bremsstrahlung and, more importantly for a low baryon density Universe such
 as ours, by
 double Compton scattering \cite{dd1982}. Electrons are also maintained at the
 equilibrium Maxwellian distribution by Compton scattering with the photons
 \cite{zl1970,ls1971}
 whose number density exceeds that of the electrons by a factor of $\sim
 10^9$. 
 Coulomb collisions efficiently maintain equilibrium between electrons and
 ions in the entire redshift range of interest to us and they can be assumed
 to have the same temperature, defined by the photon spectrum\cite{zl1970,ls1971}, Eq. \ref{te}. 

At redshifts $z\lesssim 2\times
 10^6$, bremsstrahlung and double Compton scattering become inefficient in
 creating photons, 
 however Compton scattering is still able to maintain kinetic equilibrium
 {(Bose-Einstein spectrum)} 
 until $z\approx 2\times 10^5$. This, therefore, defines  the era where it is possible to
 create $\mu$-type distortions with the $\mu$ parameter related to the 
 fractional energy $Q$ ($\equiv \Delta E/E_{\gamma}$, where
$\Delta E$ is the energy density going into the spectral distortions and
$E_{\gamma}=\aR T^4$ is the energy density of radiation and $\aR$ is the
radiation constant) injected into the radiation by\cite{sz1970,is1975b} $\mu=1.4 Q$. To  calculate the $\mu$-type
 distortions it is necessary to calculate  precisely  the suppression of $\mu$ behind
 the blackbody surface at $z\approx 2\times 10^6$. This is of course
 possible with the numerical codes such as
 CosmoTherm\footnote{www.chluba.de/CosmoTherm} \cite{cs2011}  and
 KYPRIX\cite{pb2009}, the former code includes precise calculation of 
 distortions arising from the energy injection due to 
 Silk damping. Sunyaev and Zeldovich \cite{sz1970}
 found an analytic solution for the suppression factor or blackbody visibility
 $e^{-\mathcal{T}(z)}$ and this solution was recently made even more
 precise \cite{ks2012}, allowing a fast  and accurate computation of $\mu$-type
 distortions.  These analytic solutions and the recipe for using them are given
 in Appendix \ref{appa}. COBE $95\%$ confidence level limit on the $\mu$
 parameter is 
 $\mu<9\times 10^{-5}$ \cite{cobe}. 

At redshifts $z\lesssim 2\times 10^5$, Compton scattering {is not
  sufficient to maintain a Bose-Einstein spectrum in the presence of energy
  injection} 
   and is  able to move the spectrum only partially towards the equilibrium
creating intermediate or $i$-type distortions at redshifts $z\gtrsim 1.5\times 10^4$. The distortions in this epoch  must be calculated numerically by
solving the Kompaneets equation \cite{k1956}. Numerical studies of
Kompaneets equation have been performed by many authors \cite{is1975b,ss1983,bdd1991,hs1993,pb2009,cs2011}.  Recently
a 
complete set of numerical solutions (or Green's functions) were calculated
in 
Ref. \cite{ks2012b} and the results are publicly available.  The recipe for
using these results to calculate the $i$-type part of spectral distortions
for a general energy injection scenario is  given in Appendix
\ref{appa}. {We refer to \cite{ks2012b} for a detailed discussion
  of the $i$-type distortions and how they can help distinguish between
  different energy injection scenarios, for example energy injection rate which
  is exponential in redshift such as particle decay,  and that which is a
  power law such as Silk damping. We should emphasize that an $i$-type
  component is inevitable for power law energy injection such as Silk
  damping (which itself is unavoidable) while a particle decay can happen entirely in the $\mu$ epoch
  leading to negligible $i$-type distortions. Thus the presence of absence
  of $i$-type distortions together with the shape of the $i$-type
  distortions is a powerful discriminant for different energy injection mechanisms.}

At redshifts $z\lesssim 1.5\times 10^4$, comptonization is minimal, and the
solution for the distortions is given  by a $y$-type
distortion \cite{zs1969}, which are also created at lower redshifts when
the CMB photons are
scattered in the clusters of galaxies by hot electrons. {The $y$-type distortions are expected to be dominated
by low redshift contributions coming from the epoch of  reionization, with
y parameter (see Eqs. \ref{dampeq},\ref{distdef}) given by  $y\sim
10^{-7}$, and warm hot intergalactic
medium  \cite{hss1994b,co1999,co2006,ns2001,tbo,bbps,lnb,ks2012b,ds2013} with $y\sim
10^{-6}-10^{-7}$}. The $y$-type distortions  are therefore hard to predict and must be
fitted as a free parameter during the Fisher matrix analysis. COBE $95\%$ limit
on the $y$ parameter is  $y<1.5\times 10^{-5}$ \cite{cobe}. \changeR{The
  low redshift contributions, since they originate and are dominated by
  the collapsed structures are naturally very inhomogeneous, unlike high
  redshift contributions before reionization which are homogeneous to high accuracy. Thus,} 
a future  experiment such as Cosmic Origins Explorer (COrE) \cite{core}
may be able to detect compact sources of $y$-type distortion and
thus help estimate and clean the low redshift average $y$-type distortion contribution.

There are spectral distortions, other than $y$, $\mu$ and $i$-type, which  are
created in the CMB, for example, recombination lines from the epoch of
recombination \cite{zks68,peebles68,d1975,rcs2006,cs2006b,rcs2008}    and in CMB
anisotropies from  resonant scattering of the CMB in metals
lines 
during reionization and later \cite{basu}, which we will not discuss here. We
refer to a recent 
 review \cite{sk2013} for a more complete discussion of the various  spectral
 distortions in the CMB.

\section{Spectral distortions from Silk damping: Observing 17 e-folds of
  inflation}
\begin{figure}
\resizebox{\hsize}{!}{\includegraphics{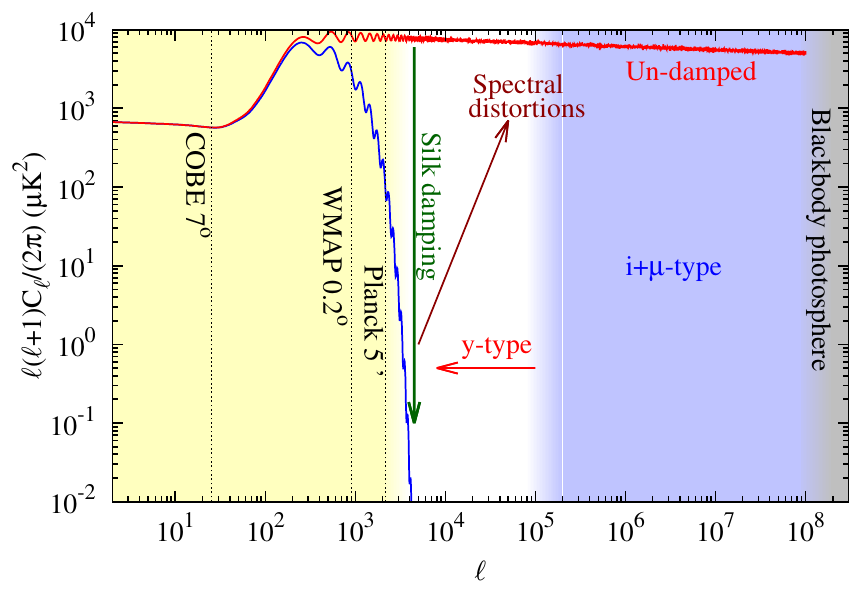}}
\caption{\label{damping}Power which disappears from the anisotropies
  appears in the monopole as spectral distortions. \changeR{CMB damped and
    undamped power spectra
were calculated using analytic approximations
\cite{jorgensen95,seljak1994,hs1995,hw1997}. Scale range probed by the CMB
anisotropy experiments such as COBE-DMR, WMAP, Planck, SPT and ACT is
marked by the shaded region on the left side of the plot. Spectral distortions probe much  smaller scales up
to
the blackbody photosphere boundary at $\ell\sim 10^8$.}}
\end{figure}
Photons diffusing through the primordial plasma erase perturbations on
small scales \cite{silk,Peebles1970,kaiser} and this effect, known as Silk
damping, has already been observed in the CMB anisotropies by the Atacama
Cosmology Telescope\cite{act} (ACT),  the South Pole Telescope\cite{spt}
(SPT), and now the Planck experiment \cite{planck} on scales up to $k\approx 0.2 \mpci$, {which is also the damping scale at $z\approx
1200$,}  where the anisotropies are
suppressed but not completely erased. \changeR{Taking into account that the largest scale we can
  observe today is the horizon scale, $k\approx 2.2\times 10^{-4}$, CMB
  anisotropies are giving us a view of inflation 
  corresponding to $\sim 6.8$ e-folds. The wavenumbers corresponding
to the photon diffusion length are $\kD\approx 1.1\times 10^4 \mpci$ (or multipole
$\ell \approx 10^8,~ 17.7$ e-folds to horizon size today)} at $z=1.95\times 10^6$
and $\kD \approx 8\mpci$ ($\ell \approx 10^5$)  at $z=1.5\times 10^4$.  On these very small
scales the anisotropies are almost completely erased from the CMB and are
therefore unobservable in the CMB anisotropy power
spectrum. 
   The energy stored in the
perturbations (or the sound waves in the primordial radiation pressure
dominated plasma) on the dissipating scales, however, does not disappear but goes into the monopole
spectrum creating $y$, $\mu$ and $i$-type distortions, see Fig. \ref{damping}. This effect was
estimated initially by Sunyaev and Zeldovich \cite{sz1970} and later by
Daly \cite{daly1991} and Hu, Scott and Silk \cite{hss94}. Recently, the
energy dissipated in Silk damping and going into the spectral distortions
was calculated precisely in \cite{cks2012}, correcting previous calculations
and also giving {a clear} physical interpretation of the effect in terms
of mixing of blackbodies \cite{cks2012,ksc2012b} \footnote{See \cite{pz2012} for a
slightly different way of calculating $\mu$-type distortions and also \cite{Weinberg1971}.}. The
calculations in \cite{cks2012} showed that photon diffusion just mixes
blackbodies and the  resulting distortion is a $y$-type distortion which
can comptonize into $i$-type or $\mu$-type distortion, depending on the
redshift. We can write down the (fractional) dissipated energy ($Q\equiv
\Delta E/E_{\gamma}$) going into the
spectral distortions as  \cite{cks2012,ksc2012b}
\begin{align}
\frac{\id Q}{\id t}&=-2\frac{\id}{\id t}\int \frac{k^2\id k}{2\pi^2}P_i^{\gamma}(k)\left[\sum_{\ell=0}^{\infty}(2\ell+1)\Theta
_{\ell}^2\right]
 \approx -2\frac{\id}{\id t}\int \frac{k^2\id k}{2\pi^2}P_i^{\gamma}(k)\left[\Theta
_{0}^2+3\Theta_1^2\right],\label{silkQ}
\end{align}

where $\Theta_{\ell}(k)$ are the spherical harmonic multipole moments of temperature
anisotropies of the CMB, $t$ is proper time and
$P_i^{\gamma}(k)=\frac{4}{0.4 R_{\nu} +1.5}P_{\zeta}\approx 1.45
P_{\zeta}$, $P_{\zeta}=(A_{\zeta}2\pi^2/k^3)(k/k_0)^{\nS -1+\frac{1}{2} \run
  (\ln k/k_0)} $,   the amplitude of 
comoving curvature perturbation  $A_{\zeta}$  is equivalent to $\Delta_R^2$ in 
Wilkinson Microwave Anisotropy Probe (WMAP) papers \cite{wmap9}, and
$R_{\nu}=\rho_{\nu}/(\rho_{\nu}+\rho_{\gamma})\approx 0.4$, $\rho_{\nu}$ is the
initial neutrino energy density and $\rho_{\gamma}$ is the initial photon
energy density, but after electron positron annihilation \cite{mabert95},
$k_0$ is the pivot point, $\nS$ is the spectral index and $\run$ is its
running \cite{mukhanov,kt1995}. The last approximate equality is valid in the tight coupling approximation
when the energy in the  multipole moments $\ell\ge 2$ can be neglected.

It is possible to evaluate the time derivative in Eq. \ref{silkQ}
explicitly using the first order Boltzmann equation for photons, yielding
an exact expression in terms of photon quadrupole, dipole and higher order
multipoles  and baryon
peculiar velocity which is valid at all times \cite{cks2012,ksc2012b}. But
in the redshift range of interest to us, $z\gtrsim 1.5\times 10^4$, the 
 tight coupling solutions are quite accurate  \cite{hs1995,dod},
 yielding the following expression for the energy injection rate \cite{ksc2012,cks2012},
\begin{align}
\frac{\id Q}{\id z}=\frac{9}{4}\frac{\id(1/\kD ^2)}{\id z}\int
\frac{\id^3k}{(2\pi)^3}P_{\gamma}^i(k)k^2e^{-2k^2/\kD ^2} \label{dqdz},
\end{align}
where $\kD$ is the damping wavenumber given by \cite{silk,Peebles1970,kaiser,weinberg},
\begin{align}
\frac{1}{\kD ^2}&=\int_z^{\infty}\id z
\frac{c(1+z)}{6H(1+R)\Ne \sigT}\left(\frac{R^2}{1+R}+\frac{16}{15}\right),\label{kd}
\end{align}
where $R\equiv \frac{3\rho_b}{4 \rho_{\gamma}}$, $\rho_b$ is the baryon
  energy density, $\sigT$ is Thomson cross section, $\me$ is the mass of electron,
$\Ne$ is the number density of electrons, $c$ is the speed of light, and
$H$ is the Hubble parameter. In general the 
Eqs. \ref{dqdz} and \ref{kd} must be evaluated numerically. However in the
radiation dominated epoch, $R\ll 1$ and for a power law initial power
spectrum, $\run=0$, we can evaluate the integrals analytically yielding,
\begin{align}
\frac{\id Q}{\id z}=\frac{3.25 A_{\zeta}}{k_0^{\nS -1}}\frac{\id(1/\kD ^2)}{\id z}2^{-(3+\nS )/2}\kD ^{\nS +1}\Gamma\left(\frac{n+1}{2}\right).
\end{align}

Once we know the energy injection rate, it is easy to calculate the
$\mu$-type and $i$-type distortions using the method given in Appendix
\ref{appa}. We should also mention that the spectral distortions from Silk
damping  may also permit us to constrain the
primordial local type non-gaussianity on these very small scales \cite{pajer2012,ganc2012}.

\section{Fisher forecast results for Pixie-like experiments}
We will now calculate the constraints we can put on the initial power
spectrum using CMB spectral distortions. Although the spectral distortions
are sensitive to cosmological parameters other than the power spectrum, as
can be seen from the equations, these sensitivities are relatively milder
and we know the other cosmological parameters to high accuracy from CMB
anisotropy and other data \cite{wmap9,spt,act,spt2,act2,bao1,bao2,bao3,h01,h02,h03,sn1,sn2}. We will therefore fix all
parameters, except for the power spectrum, to the following
values for flat $\Lambda$CDM cosmology\cite{wmap9}:  CMB temperature $\TCMB=2.725~{\rm K}$, baryon density
parameter $\Ob=0.0458$, cold dark matter $\Ocdm=0.229$, Hubble constant
$h_0=0.702$, helium fraction $0.24$, effective number of
neutrinos\cite{mm2005} $3.046$. 

We can write down the spectral distortion of CMB,  $\Delta I_{\nu}$, which
will be measured by Pixie \cite{pixie} as
\begin{align}
\Delta I_{\nu}=t I^t_{\nu}+y I^y_{\nu}+ I_{\nu}^{\rm
  damping}(\nS,A_{\zeta},\run).\label{dampeq}
\end{align}
The first term is the uncertainty in the temperature of the blackbody part
of the spectrum which is not known a priori and must be fit simultaneously
with the spectral distortions, second term is the $y$-type distortion which
has contributions from low redshift intergalactic medium and is therefore a
free parameter. The last term is the $i$-type + $\mu$-type distortions from
the dissipation of acoustic modes and is a function of the power spectrum
which we parmeterize by the amplitude, the spectral index and its
running. The formulae for different terms in Eq. \ref{dampeq} are given in Appendix \ref{appa}. If
there is new physics injecting  energy, there will be additional terms
added to the above distortion.  For example, if there is decay of particles
before $z=2\times 10^5$, it will create additional $\mu$-type distortions
and a term $\mu I_{\nu}^{\mu}$ should be fitted  with $\mu$ as a free
parameter. Adding additional parameters will of course degrade the
constraints on the primordial power spectrum and we will discuss it briefly
below. We should also include cooling of CMB due to energy
transfer to baryons which cool faster than radiation
\cite{zks68,peebles68,cs2011,ksc2012}, however, it depends only on
cosmological parameters which we have assumed to be constant and therefore
does not affect the Fisher matrix. The cooling must be included in a
precise calculation using, for example, Markov chain Monte Carlo to explore
the likelihood.

If $\Delta \nu$ is the spectral resolution of the experiment and $\delta
I(\nu)$ is the noise in each channel, the Fisher matrix is given by (e.g. 
\cite{tegmark,matsubara,detf}),
\begin{align}
F_{\alpha \beta}=\sum_j\frac{1}{\delta I(\nu_j)^2}\frac{\partial \Delta I_{\nu}}{\partial \theta_{\alpha}}(\nu_j)\frac{\partial \Delta I_{\nu}}{\partial \theta_{\beta}}(\nu_j),
\end{align}
where $\nu_j$ is the center frequency of each channel, $\theta_{\alpha,\beta} \in
(t,y,\nS,A_{\zeta},\run)$, and the sum is over all frequency channels which we
take to be from $30~{\rm GHz}$ to $500~{\rm GHz}$. The upper limit of the
usable frequency range will depend on how well we can remove the
foregrounds which become more problematic at higher
frequencies. \changeR{The residual foregrounds would need to be jointly fitted and
  marginalized over in the data analysis \cite{cobe,sb2002}.} In the
analysis below we will assume that the foregrounds have been removed at
required precision ($\sim \delta I$) using the frequency channels greater
than $500~\rm{GHz}$ which is the plan for the Pixie experiment
\cite{pixie}. In the  Pixie proposal $\Delta \nu=15 ~{\rm
  GHz}$, with a total of 400 frequency channels extending to $6 ~{\rm THz}$
and the hope is that the large number of channels at high frequencies
would help nail down the foregrounds to the desired accuracy. 

It is extremely important to study  the impact the 
foregrounds have on the ability of Pixie-like experiments to detect   spectral
distortion  but which is beyond the scope of present work. This
paper is 
only a first step in  quantifying the information content and the
detectability of the
spectral distortions with respect to constraining the primordial power
spectrum on scales $8~ \mpci \lesssim k \lesssim 10^4~\mpci$. 

The covariance matrix is just the inverse of the Fisher matrix,
$cov_{\alpha \beta}=\left[F^{-1}\right]_{\alpha \beta}$
If we are interested in only first $n$ parameters of the parameter vector
$\theta$, we can marginalize over the rest of the parameters by writing the
Fisher matrix as 
\begin{align}
\left(
\begin{array}{lr}
A & B \\
B^T & C
\end{array}
\right),
\end{align}
where the sub-matrix $A$ spans over the parameters we are interested in. The
marginalized Fisher matrix is then given by $\bar{F}=A-BC^{-1}B^T$. It can
also be obtained (if the parameters are non-degenerate) by taking the
 rows and columns  of the 
parameters we are interested in from the
covariance matrix and inverting the resulting sub-matrix. If
we want to fix some parameter at a particular value instead of
marginalizing over it, we just eliminate the row and column for that
parameter from the Fisher matrix. The marginalized
$\nu-\sigma$ ellipsoids are then given by $\Delta \theta^T
\bar{F} \Delta \theta =\chi^2(\nu)$ where $\Delta \theta=\theta-\theta_{\rm fiducial}$
and the Fisher matrix is also evaluated at the fiducial values of the
parameters. For two parameters, $\chi^2(1)=2.3, \chi^2(2)=6.17$.

\begin{figure}
\resizebox{\hsize}{!}{\includegraphics{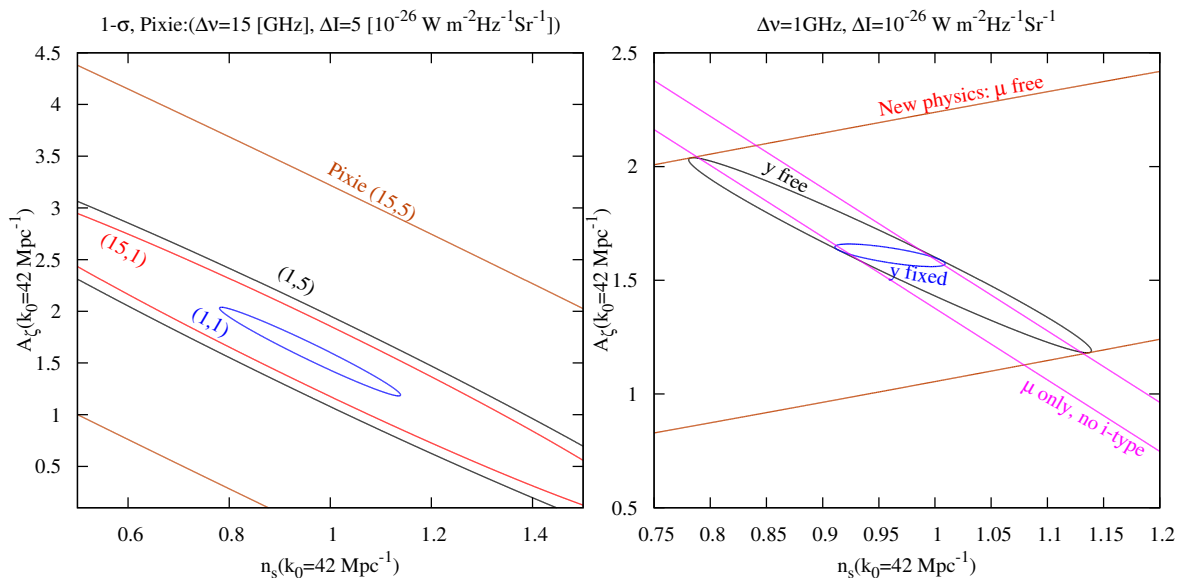}}
\caption{\label{Fisher1} Marginalized Fisher matrix constraints on
  amplitude and spectral index of primordial power spectrum defined with
  respect to the pivot point $k_0=42~\mpci$. Left panel shows constraints
  for different spectral resolutions and sensitivities of Pixie-like
  experiments. The labels for different contours are resolution in units of
  GHz and $\delta I$ in units \changeR{of ${\rm 10^{-26}W m^{-2}Sr^{-1}Hz^{-1}}$,
  ($\Delta \nu, \delta I$).} Right panel demonstrates degeneracies between
  different parameters for resolution $1~{\rm GHz}$ and \changeR{sensitivity ${10^{-26}\rm W
    m^{-2}Sr^{-1}Hz^{-1}}$.} The curve labeled $y$ free is the normal
  contour we expect marginalized over $y$ and $t$. If we ignore the
  information in the shape of the $i$-type distortions and consider only
  $\mu$-type distortions, as in studies so far, we get the curve labeled
  '$\mu$ only'. If we add an additional free parameter, $\mu$, which may
  arise from new physics such as decay of particles, we get the
  '$\mu$-free' curve. The curve labeled '$y$-fixed' is the one we get if we
  assume we can predict and fix the $y$-type distortions from  low redshifts.
   }
\end{figure}

Let us first consider the constraints we can obtain from spectral
distortions alone. For this we take the pivot point $k_0=42 ~\mpci$, which
is approximately in the middle of the $i$-type distortions epoch, and
constrain the amplitude and the spectral index on small scales (without
running and large scale power spectrum constraints). We thus have the
parameter vector $\theta=(\nS(k_0=42~\mpci), A_{\zeta}(k_0=42~\mpci),
t,y)$. We take the fiducial model $\nS(k_0=42~\mpci)=0.96,
A_{\zeta}(k_0=42~\mpci)=1.61\times 10^{-9}$ and marginalize over $t$ and
$y$. The resulting $68\%$ contours are shown in Fig. \ref{Fisher1} for
Pixie \cite{pixie}
spectral resolution $\delta \nu=15{\rm GHz}$ and sensitivity $\delta
I(\nu)=5\times 10^{-26} {\rm W m^{-2}Sr^{-1}Hz^{-1}}$. We also show the
contours obtained by increasing the spectral resolution to $1~{\rm GHz}$
or/and sensitivity to $\delta
I(\nu)=10^{-26} {\rm W m^{-2}Sr^{-1}Hz^{-1}}$. We should emphasize that
these are completely independent constraints on the power spectrum on
scales $8~\mpci \lesssim k \lesssim 10^4~\mpci$, where there are currently
absolutely no constraints. Thus, although they may look much weaker compared
to the constraints from CMB anisotropies \cite{wmap9}, they still represent
significant improvement in our knowledge  of the cosmological initial conditions.

The right panel in Fig. \ref{Fisher1} demonstrates the degeneracies between
different types of distortions. The curve marked  '$y$ free' is the
same as the (1,1) curve in the left panel, marginalized over $y$ and
$t$. The curve labeled '$\mu$ only, no $i$-type' is the one obtained if we
ignore that the $i$-type distortions have a  characteristic shape with
information about the spectral index and divide all the energy released
into $y$-type ($z\lesssim 5\times 10^4$) and $\mu$-type ($z\gtrsim 5\times
10^4$) distortions, as in previous studies
(c.f. Fig. 17 in \cite{cks2012}). Clearly it is not possible to constrain two parameters with
  one observable, $\mu$, in this case and our two parameters are
  completely degenerate. Inclusion of the  $i$-type distortions gives the '$y$
  free' curve and converts the straight line contours into an ellipse
  demonstrating the information contained in the shape of the $i$-type
  distortions. A different way of seeing this  is by making the $\mu$
  parameter free, for example, as a result of new physics such as decaying
  particles at $z\gtrsim 2\times 10^5$. Note that we already have $y$ and
  $t$ free, so the constraints in the '$\mu$ free'  contours are coming solely from the shape of the
  $i$-type distortions. The fact that we can get any constraints at all in
  this last case demonstrates that the $i$-type distortions are not completely
  degenerate with and cannot be mimicked by a combination of $y$, $\mu$
  distortions and $t$ parameter. Finally the curve labeled '$y$ fixed' is
  obtained by eliminating the row and column corresponding to $y$ parameter
  from the Fisher matrix, thus fixing $y$. Comparing with '$y$ free'
  curve, this tells us how the presence of stars and galaxies (responsible
  for reionization, WHIM which give low redshift $y$-distortions)  in the
  Universe limits our ability to measure the primordial power spectrum with
  the CMB 
  spectral distortions.

\begin{figure}
\resizebox{\hsize}{!}{\includegraphics{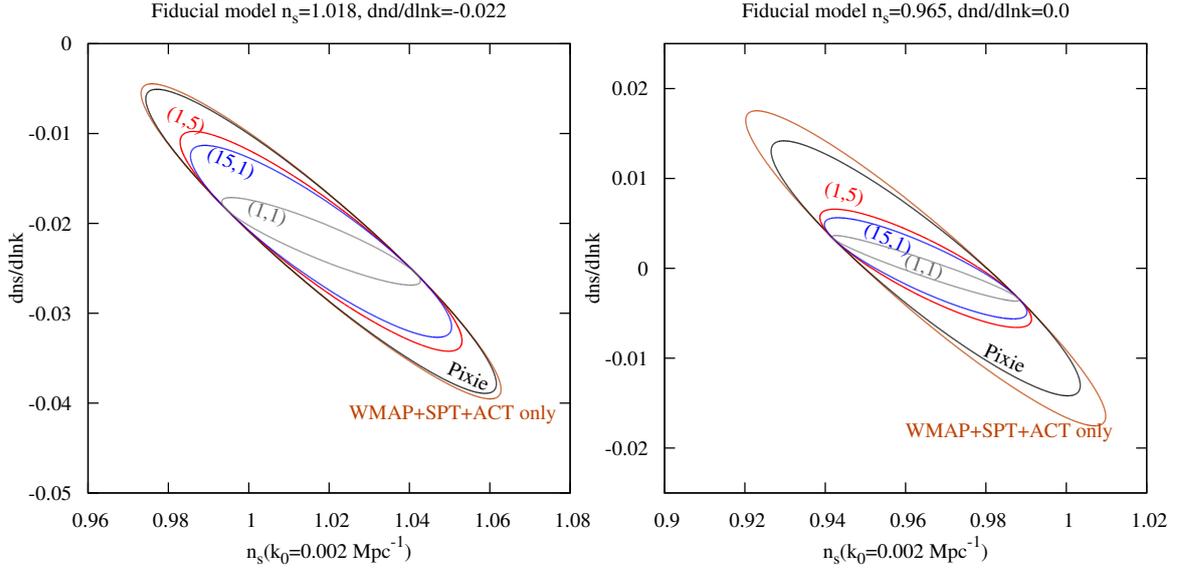}}
\caption{\label{Fisherwmap}  Constraints ($68\%$ confidence level) by combining WMAP+SPT+ACT \cite{wmap9,spt,spt2,act,act2}
  with spectral distortions from  Pixie-like experiments for two different
  fiducial models. Labels are same as
in Fig. \ref{Fisher1}. A fiducial model with zero running has more power on
small scales and therefore gives tighter constraints.}
\end{figure}

The model considered above is perhaps the most general model we can hope to
constrain with the spectral distortions. A very  restrictive  model on the
other hand is a model with running spectral index, which applies to both
the large scale anisotropies and the spectral distortion. For this model, we
take the pivot point at $k=0.002~\mpci$ as in the WMAP papers. So we can use
the Markov chains provided by the WMAP team and combine the CMB anisotropy
data with the spectral distortions to see how the spectral distortions improve
the constraints on the spectral index and its running. We use the Markov chain
with the running spectral index and including ACT \cite{act,act2} and SPT
\cite{spt,spt2} data provided by the WMAP team\footnote{\url{http://lambda.gsfc.nasa.gov/product/map/current/}} \cite{wmap9} and use CosmoMC
\cite{cosmomc} to extract the covariance matrix for $\nS, \run, A_{\zeta}$
from it. We use two fiducial models, one with best fit WMAP values for the
running model $\nS=1.018, \run=-0.022, 10^9A_{\zeta}=2.345$ and a second
one 
with  $\nS=0.965, \run=0, 10^9A_{\zeta}=2.43$ to investigate how the change in
fiducial model affects the constraints. We  marginalize over $y$, $t$ and $A_{\zeta}$ and give the
$1-\sigma$ contours for $\nS,\run$ for different spectral resolution and
sensitivities for the Pixie-like experiments in Fig. \ref{Fisherwmap}.
It is clear from this figure that spectral distortion detection would be
able to improve the constraints in this simple but restrictive model. The
constraints are sensitive to the fiducial model. This is expected since a
fiducial model with negative running will give much smaller distortions compared
to a model with zero running and the effect is amplified because of the long separation of scales
between the pivot point a $k=0.002~\mpci$ and the relevant damping scales
at $k\gtrsim 8~\mpci$. 

\begin{figure}
\resizebox{\hsize}{!}{\includegraphics{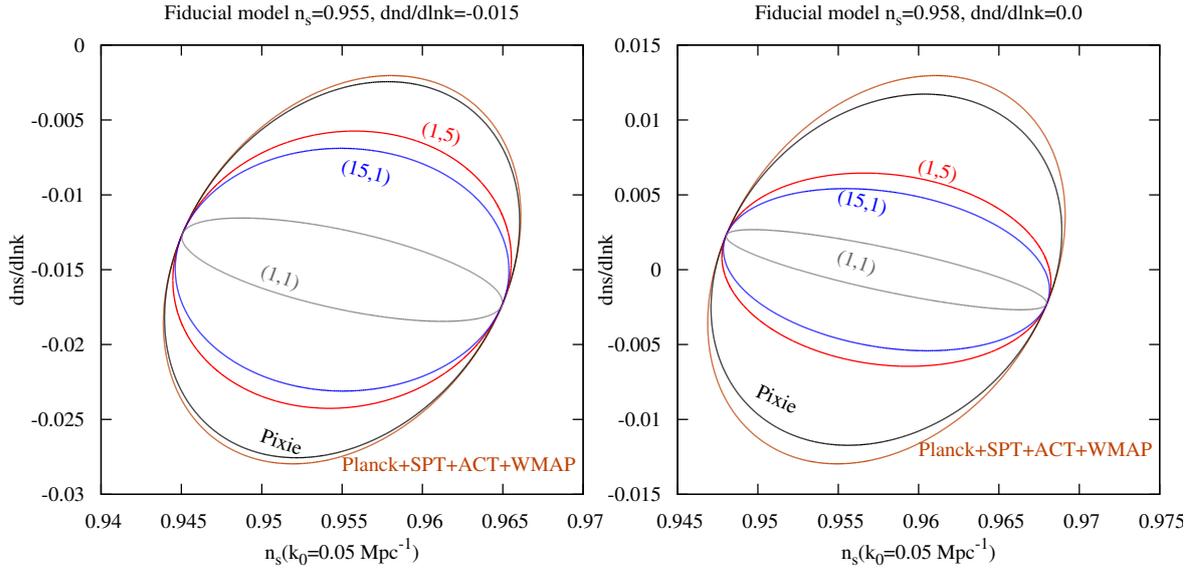}}
\caption{\label{Fisherplanck} Constraints ($68\%$ confidence level)
  obtained by combining the Planck+SPT+ACT+WMAP polarization
  \cite{planck,wmap9,spt3,act3} with spectral distortions from Pixie-like
  experiments. Labels are same as in Fig. \ref{Fisher1}. Planck mean fit 
  parameters are used for the fiducial model in the left panel of the
  plot. The right panel uses a fiducial model with zero running leading to
  stronger spectral distortions and tighter constraints.}
\end{figure}

Planck CMB  experiment's cosmology results\footnote{Based on observations obtained with Planck (\url{http://www.esa.int/Planck}), an ESA science mission with instruments and contributions directly funded by ESA Member States, NASA, and Canada.} are now available  \cite{planck}. To estimate how an improvement in the large scale
constraints on the primordial power spectrum affect the information coming from
spectral distortions, we use the covariance matrix for baseline
$\Lambda$CDM + running
model provided by the Planck
team\cite{planck} which in addition uses  polarization data  from WMAP
\cite{wmap} and high $\ell$ data from SPT \cite{spt3} and ACT
\cite{act3}. For combining Planck results with spectral distortions, we use
the Planck mean fit  parameters, $\Ob h^2=0.02225, \Ocdm h^2=0.1205,
h=0.672$  and fiducial model with $\nS=0.955, \run=-0.015, \ln(10^{10}
A_{\zeta})=3.12, k_0=0.05~\mpci$. The results are plotted in
Fig. \ref{Fisherplanck}. We again demonstrate the effect of fiducial model
in the right panel of Fig \ref{Fisherplanck} where the fiducial model is
taken to be  $\nS=0.958, \run=0, \ln(10^{10}
A_{\zeta})=3.1, k_0=0.05~\mpci$ but the same Planck covariance matrix and
other parameters as the
left panel.  By
comparing Figs. \ref{Fisherwmap} and \ref{Fisherplanck} it is clear that
there is additional and complementary information coming from the spectral
distortions irrespective of the improvements in the large scale constraints
from the CMB anisotropies. 

\section{Concluding remarks}
We have calculated the Fisher matrix forecasts for the possible constraints
on the primordial power spectrum using CMB spectral distortions arising from
Silk damping. These results, of course, come with the usual caveats
associated with the Fisher matrix analysis. In particular, the spectral
distortions are a non-linear function of the primordial power spectrum
parameters and are most likely non-Gaussian. In addition, adiabatic cooling
of baryons \cite{cs2011,ksc2012} reduces the signal and determines the smallest positive
distortions that can be detected.  Our analysis does not take this into
account, since the constant terms drop out in the derivatives used to
calculate Fisher matrices. {For the cosmological parameters preferred
by Planck, the adiabatic cooling dominates  over the Silk damping for
$\run<-0.08$, for the total $i$-type $+$ $\mu$-type distortion, which is
ruled out at high significance and our
Fisher matrix calculations are a good approximation within the allowed
region.}  Our calculation represents the
first step in quantifying the information stored in the spectral
distortions, in particular the $i$-type distortions and paves the way for a  more careful analysis using Markov chain Monte Carlo
techniques in the future. Fisher matrix analysis, in particular, shows approximately how much
improvement in constraints can be expected if the spectral resolution or the sensitivity
of Pixie is made better and also shows how these constraints are sensitive
to  the choice of  fiducial models. \changeR{We should point out that since
  we are looking at broad features, which are already resolved at $15 {\rm
    GHz}$ resolution, the  improvement in just the frequency resolution but
  keeping the total sensitivity summed over all channels constant does not
  help. However by adding more channels but maintaining  the same sensitivity in \emph{each}
  channel  improves the constraints just because we have more data. Thus
  the contours marked $(1,5)$ add $15$ times more channels and are
  equivalent to an improvement in sensitivity of $\sqrt{15}\sim 4$ over
  Pixie fiducial proposal. The Pixie contours assume the
  sensitivity achieved for a 4-year mission specified in the Pixie proposal
  \cite{pixie} when $30\%$ of the time is used in doing absolute
  measurements of the CMB spectrum. An improvement in sensitivity of factor
  of $2-3$ can be  achieved by observing for more time ($5-11$
  years). Additional improvement may be possible by using more
  detectors. The constraints in the curves marked $(1,5)$ and $(15,1)$
  correspond to a $4$ and $5$ times 
   more sensitivity respectively compared to Pixie  proposal and are thus
   in principle achievable. Even the curves labeled $(1,1)$, corresponding to $20$ times
   more sensitivity compared to Pixie, are possible with present technology
   \cite{fm2002,pixie,core}.}

We have also shown that there is important information in the shape of the
spectral distortions coming from the $i$-type distortions, which has so far been ignored in constraint
calculations, although this information  was available from the numerical computations of the spectral
distortions. {In particular, the  $i$-type distortions are very important in
breaking the degeneracy between the amplitude and the spectral index of the
primordial power spectrum on the scales $8\lesssim k \lesssim 10^4 ~\mpci$.} More important than the error bars on different power spectrum
parameters is the fact that with spectral distortions we will be able to
extend our knowledge of initial conditions to completely new scales
separated by many orders of magnitude from the information available from
CMB anisotropy and large scale structure data. In inflationary terms, this
amounts to extending our view of inflation from $\sim 6-7$ e-folds at present
to $\sim 17$ e-folds, which might well be a significant fraction of the full
inflationary epoch. 

\begin{appendix}
\section{Recipe for the calculation of CMB spectral distortions for
  a general energy injection scenario}\label{appa}
To calculate the spectral distortions arising from a given energy injection
rate, $dQ/dz$, where $Q=\Delta E/E_{\gamma}$ is the fractional energy
injected into the CMB, we must solve Kompaneets equation\cite{k1956}  including photon
production and absorption due to bremsstrahlung and double Compton
scattering \cite{sz1970}.  The problem is however complicated by the fact
that the electron
temperature  enters the partial differential equation describing the
evolution of photon intensity $I_{\nu}$, or equivalently the photon
occupation number $n(x)=c^2/(2 h \nu^3) I_{\nu}$, itself depends on the
photon spectrum \cite{zl1970,ls1971}
\begin{align}
\frac{T_e}{T}=\frac{\int(n+n^2)x^4dx}{4\int n x^3 dx}\label{te}
\end{align}

At high redshifts, when Compton scattering is able to establish
Bose-Einstein spectrum, an analytic solution for the evolution of $\mu$
parameter accurate to $\sim 1\%$ can be found and is given by
\cite{sz1970,ks2012}
\begin{align}
\mu=1.4\int_{z_{\rm max}}^{z_{\mu}}\id
z\left( \frac{\SE}{\SQ}-\frac{\SE}{\SQ}^{\rm cooling}\right)e^{-\mT(z)}\label{musol}
\end{align}
where 
\begin{align}
\mT(z)&\approx
1.007\left[\left(\frac{1+z}{1+\zdC}\right)^5+\left(\frac{1+z}{1+\zbr}\right)^{5/2}\right]^{1/2}+1.007
\epsilon
\ln\left[\left(\frac{1+z}{1+z_{\epsilon}}\right)^{5/4}+\sqrt{1+\left(\frac{1+z}{1+z_{\epsilon}}\right)^{5/2}}\right]\nonumber\\
&+\left[\left(\frac{1+z}{1+\zdC'}\right)^3+\left(\frac{1+z}{1+\zbr'}\right)^{1/2}\right],\label{optnew}
\end{align}

\begin{align}
\zdC&=\left[\frac{25 \Or H(0)^2}{4 C^2\aC\adC}\right]^{1/5},
&\zbr&=\left[\frac{25 \Or H(0)^2}{4 C^2\aC\abr}\right]^{2/5}&\nonumber\\
z_{\epsilon}&=\left[\frac{\abr}{\adC}\right]^{2/5},
&\epsilon&=\left[\frac{4 C^2\abr^2\aC}{25 \adC\Or H(0)^2}\right]^{1/2},&\\
\zdC'&=\left[\frac{3 \Or^{1/2} H(0)}{2.958 C\adC}\right]^{1/3},
&\zbr'&=\left[\frac{ \Or^{1/2} H(0)}{5.916 C\abr}\right]^{2}&\nonumber\\
\aC&=\Ne_0\sigT c\frac{\kB \TCMB}{\me c^2},
&\adC&=\Ne_0\sigT c \frac{4\alphafs}{3\pi} \left(\frac{\kB \TCMB}{\me
    c^2}\right)^2 \gdC(\xe)I_{\rm dC}&\nonumber\\
\abr&=\Ne_0\sigT c\frac{\alphafs \nB_0}{(24\pi^3)^{1/2}}\left(\frac{\kB
\TCMB}{\me c^2}\right)^{-7/2}\left(\frac{h}{\me
c}\right)^3\gbr(\xe,\Te),\nonumber
\end{align}
 $C=0.7768$, $I_{\rm dC}\approx 25.976$, $\gdC=1.005$ and
$\gbr=2.99$, all quantities with subscript zero are evaluated at redshift
$z=0$, $\alphafs$ is the fine structure constant and $\Or$ is the total radiation energy density parameter with  
relativistic neutrinos. The above equations assume that there is injection
of  only energy. If
there is also significant injection  of photons (other than  bremsstrahlung and double
Compton) then an additional term, $-2.404 \frac{\id N}{\id z}$ must be
added in the brackets in Eq. \ref{musol}, where $N$ is the fractional
change in the number density of photons. $z_{\rm max}$ should be taken to
be sufficiently behind blackbody surface at $\zdC\approx 1.96\times 10^6$,
$z_{\rm max}=5\times 10^6$ should be sufficient for most energy injection
scenarios. $z_{\mu}$ is the boundary of transition from $\mu$-type to
$i$-type distortions and is discussed below. We have also accounted for the
cooling of radiation because of energy transfer to baryons which cool
faster than radiation with the expansion of the Universe and which has a simple
expression before the start of the recombination
\cite{zks68,peebles68,cs2011,ksc2012},
\begin{align}
\frac{\SE}{\SQ}^{\rm cooling}=\frac{3}{2}\frac{\kB  (\nH+\nHe+\Ne)}{\aR
  T^3 (1+z)},
\end{align}
where $\nH$ and $\nHe$ are the number densities of hydrogen and helium
nuclei.

For the $i$-type distortions we must solve the Kompaneets equation numerically.
It turns out that 
the Kompaneets equation can be cast entirely in terms of dimensionless
variables using $\yg$ as the  time variable, 
\begin{align}
\yg(z,z_{\rm{inj}})=-\int_{z_{\rm{inj}}}^{z}dz\frac{k_B\sigma_T}{m_e
  c}\frac{n_eT}{H(1+z)},\label{yz}
\end{align}
Solving Kompaneets equation with the initial spectrum corresponding to a
$y$-type distortion, it was found\cite{ks2012b} that the spectrum starts deviating from
$y$-type distortion at $1\%$ level at $\yg=0.01$ and is with $1\%$ of a
$\mu$-type distortion at $\yg=2$.  We thus define the boundaries of the $i$-type
epoch by $\yg(0,z_{\mu})=2$ and $\yg(0,z_{y})=0.01$. At $z>z_{\mu}\approx
2\times 10^5$ we have
$\mu$-type distortions and at $z<z_y\approx 1.5\times 10^4$ we have
$y$-type distortions. 

The total distortion, excluding $y$-type, is now given by,
\begin{align}
\Delta I_{\nu}=\frac{2 h \nu^3}{c^2}\left[\sum_i
  \frac{n_i}{Q_{\rm ref}}\left(\frac{\SE}{\SQ}-\frac{\SE}{\SQ}^{\rm
      cooling}\right)\frac{\id z}{\id \yg}\delta 
  \yg_i +\mu n_{\mu}\right],
\end{align}
where all terms are evaluated at redshift $z_{\rm inj}^i$ related to $\yg_i$ by
$\yg_i=\yg(0,z_{\rm
  inj}^i)$, the sum is over values of $\yg$ finely sampled between $0.01$ and
$2$, $n_i(\yg_i)$ is the intermediate spectrum obtained by evolving
$y$-type distortion with energy $Q_{\rm ref}$ from $\yg=0$ to $\yg_i$ with
Kompaneets equation,
$\delta \yg_i=(\yg_{i+1}-\yg_{i-1})/2$. The $i-$type spectra, $n_i$, sampled at intervals $\delta
\yg_i=0.001$ for $\yg<1$ and $\delta
\yg_i=0.01$ for $1<\yg<10$ for $Q_{\rm ref}=4\times 10^{-5}$ are available
at \url{http://www.mpa-garching.mpg.de/~khatri/idistort.html} along with a
Mathematica code which implements the above recipe. The above formula
simply calculates the energy injected in each small redshift interval and
adds the appropriate distortion to the total. With the redshift/$\yg$ bins
defined above the accuracy of the final spectrum is $\sim 1\%$. 
The most time consuming part of the above calculation is the calculation of
$\yg(0,z_{\rm inj})$ as a function of $z_{\rm inj}$, but it can be stored
and reused if
the  normal cosmological
parameters are not changing significantly. 

Finally, $y$-type , $\mu$-type and $t$-type
occupation numbers are\footnote{{These definitions are with respect to a
  reference blackbody which is defined by the photon number density, see
  \cite{ksc2012b,ks2012b} for details.}} \cite{zs1969,sz1970,is1975b}
\begin{align}
n_y&=\frac{xe^x}{(e^x-1)^2}\left[x\left(\frac{e^x+1}{e^x-1}\right)-4\right]\nonumber\\
n_{\mu}&=\frac{\mu e^x}{\left(e^{x}-1\right)^2}\left(\frac{x}{2.19}-1\right)\label{distdef}\\
n_t&=\frac{x e^x}{\left(e^{x}-1\right)^2}\nonumber
\end{align}

\changeR{The recipe listed above is accurate to better than a $1\%$ at
  $x\gtrsim 0.4-0.5, \nu\gtrsim 25$. The corrections because of
  the cooling the baryons start becoming important and dominating at low frequencies
  ($x\lesssim 0.1, \nu\lesssim 5{\rm GHz}$) where
  bremsstrahlung tries to bring CMB spectrum in equilibrium with the
  slightly colder
  baryons \cite{cs2011}.  This effect is  not accounted for in the above
  recipe. For proposals like Pixie, which cover a frequency range of
  $\ge 30 {\rm GHz}$, the above algorithm is therefore adequate.}
\end{appendix}
\bibliographystyle{JHEP}
\bibliography{fisher}

\providecommand{\href}[2]{#2}\begingroup\raggedright\begin{thebibliography}{10}

\bibitem{cobe}
D.~J. {Fixsen}, E.~S. {Cheng}, J.~M. {Gales}, J.~C. {Mather}, R.~A. {Shafer},
  and E.~L. {Wright}, {\it {The Cosmic Microwave Background Spectrum from the
  Full COBE FIRAS Data Set}},  {\em \apj} {\bf 473} (1996) 576.

\bibitem{sz1970}
R.~A. {Sunyaev} and Y.~B. {Zeldovich}, {\it {The interaction of matter and
  radiation in the hot model of the Universe, II}},  {\em \apss} {\bf 7} (1970)
  20--30.

\bibitem{dd1982}
L.~{Danese} and G.~{de Zotti}, {\it {Double Compton process and the spectrum of
  the microwave background}},  {\em \aap} {\bf 107} (1982) 39--42.

\bibitem{zl1970}
Y.~B. {Zeldovich} and E.~V. {Levich}, {\it {Stationary state of electrons in a
  non-equilibrium radiation field.}},  {\em Soviet Journal of Experimental and
  Theoretical Physics Letters} {\bf 11} (1970) 35--38.

\bibitem{ls1971}
E.~V. {Levich} and R.~A. {Sunyaev}, {\it {Heating of Gas near Quasars,
  Seyfert-Galaxy Nuclei, and Pulsars by Low-Frequency Radiation.}},  {\em
  \sovast} {\bf 15} (1971) 363.

\bibitem{is1975b}
A.~F. {Illarionov} and R.~A. {Sunyaev}, {\it {Comptonization, characteristic
  radiation spectra, and thermal balance of low-density plasma}},  {\em
  \sovast} {\bf 18} (1975) 413--419.

\bibitem{cs2011}
J.~{Chluba} and R.~A. {Sunyaev}, {\it {The evolution of CMB spectral
  distortions in the early Universe}},  {\em \mnras} {\bf 419} (2012)
  1294--1314.

\bibitem{pb2009}
P.~{Procopio} and C.~{Burigana}, {\it {A numerical code for the solution of the
  Kompaneets equation in cosmological context}},  {\em \aap} {\bf 507} (2009)
  1243--1256.

\bibitem{ks2012}
R.~{Khatri} and R.~A. {Sunyaev}, {\it {Creation of the CMB spectrum: precise
  analytic solutions for the blackbody photosphere}},  {\em \jcap} {\bf 6}
  (2012) 38.

\bibitem{k1956}
A.~S. {Kompaneets}, {\it {The establishment of thermal equilibrium between
  quanta and electrons}},  {\em Zh. Eksp. Teor. Fiz.} {\bf 31} (1956) 876--875.

\bibitem{ss1983}
J.~{Silk} and A.~{Stebbins}, {\it {Decay of long-lived particles in the early
  universe}},  {\em \apj} {\bf 269} (June, 1983) 1--12.

\bibitem{bdd1991}
C.~{Burigana}, L.~{Danese}, and G.~{de Zotti}, {\it {Formation and evolution of
  early distortions of the microwave background spectrum - A numerical study}},
   {\em \aap} {\bf 246} (1991) 49--58.

\bibitem{hs1993}
W.~{Hu} and J.~{Silk}, {\it {Thermalization and spectral distortions of the
  cosmic background radiation}},  {\em \prd} {\bf 48} (1993) 485--502.

\bibitem{ks2012b}
R.~{Khatri} and R.~A. {Sunyaev}, {\it {Beyond y and {$\mu$}: the shape of the
  CMB spectral distortions in the intermediate epoch, $1.5 {\times} 10^{4}
  \lesssim z \lesssim 2 {\times} 10^{5}$}},  {\em \jcap} {\bf 9} (2012) 16.

\bibitem{zs1969}
Y.~B. {Zeldovich} and R.~A. {Sunyaev}, {\it {The Interaction of Matter and
  Radiation in a Hot-Model Universe}},  {\em \apss} {\bf 4} (1969) 301--316.

\bibitem{hss1994b}
W.~{Hu}, D.~{Scott}, and J.~{Silk}, {\it {Reionization and cosmic microwave
  background distortions: A complete treatment of second-order Compton
  scattering}},  {\em \prd} {\bf 49} (1994) 648--670.

\bibitem{co1999}
R.~{Cen} and J.~P. {Ostriker}, {\it {Where Are the Baryons?}},  {\em \apj} {\bf
  514} (1999) 1--6.

\bibitem{co2006}
R.~{Cen} and J.~P. {Ostriker}, {\it {Where Are the Baryons? II. Feedback
  Effects}},  {\em \apj} {\bf 650} (2006) 560--572.

\bibitem{ns2001}
B.~B. {Nath} and J.~{Silk}, {\it {Heating of the intergalactic medium as a
  result of structure formation}},  {\em \mnras} {\bf 327} (2001) L5--L9.

\bibitem{tbo}
H.~{Trac}, P.~{Bode}, and J.~P. {Ostriker}, {\it {Templates for the
  Sunyaev-Zel'dovich Angular Power Spectrum}},  {\em \apj} {\bf 727} (2011) 94,
  [\href{http://xxx.lanl.gov/abs/1006.2828}{{\tt arXiv:1006.2828}}].

\bibitem{bbps}
N.~{Battaglia}, J.~R. {Bond}, C.~{Pfrommer}, and J.~L. {Sievers}, {\it {On the
  Cluster Physics of Sunyaev-Zel'dovich and X-Ray Surveys. II. Deconstructing
  the Thermal SZ Power Spectrum}},  {\em \apj} {\bf 758} (2012) 75,
  [\href{http://xxx.lanl.gov/abs/1109.3711}{{\tt arXiv:1109.3711}}].

\bibitem{lnb}
L.~D. {Shaw}, D.~{Nagai}, S.~{Bhattacharya}, and E.~T. {Lau}, {\it {Impact of
  Cluster Physics on the Sunyaev-Zel'dovich Power Spectrum}},  {\em \apj} {\bf
  725} (2010) 1452--1465, [\href{http://xxx.lanl.gov/abs/1006.1945}{{\tt
  arXiv:1006.1945}}].

\bibitem{ds2013}
K.~{Dolag} and R.~{Sunyaev}, {\it {Relative velocity of dark matter and barions
  in clusters of galaxies and measurements of their peculiar velocities}},
  {\em ArXiv e-prints} (2013) [\href{http://xxx.lanl.gov/abs/1301.0024}{{\tt
  arXiv:1301.0024}}].

\bibitem{core}
{The COrE Collaboration}, {\it {COrE (Cosmic Origins Explorer) A White Paper}},
   {\em ArXiv e-prints} (2011) [\href{http://xxx.lanl.gov/abs/1102.2181}{{\tt
  arXiv:1102.2181}}].

\bibitem{zks68}
Y.~B. {Zeldovich}, V.~G. {Kurt}, and R.~A. {Sunyaev}, {\it {Recombination of
  Hydrogen in the Hot Model of the Universe}},  {\em Zh. Eksp. Teor. Fiz.} {\bf
  55} (1968) 278.

\bibitem{peebles68}
P.~J.~E. {Peebles}, {\it {Recombination of the Primeval Plasma}},  {\em \apj}
  {\bf 153} (1968) 1.

\bibitem{d1975}
V.~K. {Dubrovich}, {\it {Hydrogen recombination lines of cosmological origin}},
   {\em Soviet Astronomy Letters} {\bf 1} (1975) 196.

\bibitem{rcs2006}
J.~A. {Rubi{\~n}o-Mart{\'{\i}}n}, J.~{Chluba}, and R.~A. {Sunyaev}, {\it {Lines
  in the cosmic microwave background spectrum from the epoch of cosmological
  hydrogen recombination}},  {\em \mnras} {\bf 371} (2006) 1939--1952.

\bibitem{cs2006b}
J.~{Chluba} and R.~A. {Sunyaev}, {\it {Free-bound emission from cosmological
  hydrogen recombination}},  {\em \aap} {\bf 458} (2006) L29--L32.

\bibitem{rcs2008}
J.~A. {Rubi{\~n}o-Mart{\'{\i}}n}, J.~{Chluba}, and R.~A. {Sunyaev}, {\it {Lines
  in the cosmic microwave background spectrum from the epoch of cosmological
  helium recombination}},  {\em \aap} {\bf 485} (2008) 377--393.

\bibitem{basu}
K.~{Basu}, C.~{Hern{\'a}ndez-Monteagudo}, and R.~A. {Sunyaev}, {\it {CMB
  observations and the production of chemical elements at the end of the dark
  ages}},  {\em \aap} {\bf 416} (2004) 447--466.

\bibitem{sk2013}
R.~A. {Sunyaev} and R.~{Khatri}, {\it {Unavoidable CMB spectral features and
  blackbody photosphere of our Universe}},  {\em ArXiv e-prints} (Feb., 2013)
  [\href{http://xxx.lanl.gov/abs/1302.6553}{{\tt arXiv:1302.6553}}].

\bibitem{jorgensen95}
H.~E. {Jorgensen}, E.~{Kotok}, P.~{Naselsky}, and I.~{Novikov}, {\it {Evidence
  for Sakharov oscillations of initial perturbations in the anisotropy of the
  cosmic microwave background}},  {\em \aap} {\bf 294} (Feb., 1995) 639--647.

\bibitem{seljak1994}
U.~{Seljak}, {\it {A two-fluid approximation for calculating the cosmic
  microwave background anisotropies}},  {\em \apjl} {\bf 435} (Nov., 1994)
  L87--L90, [\href{http://xxx.lanl.gov/abs/astro-ph/9406050}{{\tt
  astro-ph/9406050}}].

\bibitem{hs1995}
W.~{Hu} and N.~{Sugiyama}, {\it {Anisotropies in the cosmic microwave
  background: an analytic approach}},  {\em \apj} {\bf 444} (May, 1995)
  489--506, [\href{http://xxx.lanl.gov/abs/astro-ph/9407093}{{\tt
  astro-ph/9407093}}].

\bibitem{hw1997}
W.~{Hu} and M.~{White}, {\it {The Damping Tail of Cosmic Microwave Background
  Anisotropies}},  {\em \apj} {\bf 479} (Apr., 1997) 568,
  [\href{http://xxx.lanl.gov/abs/astro-ph/9609079}{{\tt astro-ph/9609079}}].

\bibitem{silk}
J.~{Silk}, {\it {Cosmic Black-Body Radiation and Galaxy Formation}},  {\em ApJ}
  {\bf 151} (1968) 459.

\bibitem{Peebles1970}
P.~J.~E. {Peebles} and J.~T. {Yu}, {\it {Primeval Adiabatic Perturbation in an
  Expanding Universe}},  {\em \apj} {\bf 162} (1970) 815.

\bibitem{kaiser}
N.~{Kaiser}, {\it {Small-angle anisotropy of the microwave background radiation
  in the adiabatic theory}},  {\em MNRAS} {\bf 202} (1983) 1169--1180.

\bibitem{act}
{R. Hlozek et al.}, {\it {The Atacama Cosmology Telescope: A Measurement of the
  Primordial Power Spectrum}},  {\em \apj} {\bf 749} (2012) 90.

\bibitem{spt}
{R. Keisler et al.}, {\it {A Measurement of the Damping Tail of the Cosmic
  Microwave Background Power Spectrum with the South Pole Telescope}},  {\em
  \apj} {\bf 743} (2011) 28.

\bibitem{planck}
{Planck Collaboration}, P.~A.~R. {Ade}, N.~{Aghanim}, C.~{Armitage-Caplan},
  M.~{Arnaud}, and et~al., {\it {Planck 2013 results. XVI. Cosmological
  parameters}},  {\em ArXiv e-prints} (Mar., 2013)
  [\href{http://xxx.lanl.gov/abs/1303.5076}{{\tt arXiv:1303.5076}}].

\bibitem{daly1991}
R.~A. {Daly}, {\it {Spectral distortions of the microwave background radiation
  resulting from the damping of pressure waves}},  {\em \apj} {\bf 371} (1991)
  14--28.

\bibitem{hss94}
W.~{Hu}, D.~{Scott}, and J.~{Silk}, {\it {Power spectrum constraints from
  spectral distortions in the cosmic microwave background}},  {\em ApJl} {\bf
  430} (1994) L5--L8.

\bibitem{cks2012}
J.~{Chluba}, R.~{Khatri}, and R.~A. {Sunyaev}, {\it {CMB at $2 {\times} 2$
  order: the dissipation of primordial acoustic waves and the observable part
  of the associated energy release}},  {\em \mnras} {\bf 425} (2012)
  1129--1169.

\bibitem{ksc2012b}
R.~{Khatri}, R.~A. {Sunyaev}, and J.~{Chluba}, {\it {Mixing of blackbodies:
  entropy production and dissipation of sound waves in the early Universe}},
  {\em \aap} {\bf 543} (2012) A136.

\bibitem{pz2012}
E.~{Pajer} and M.~{Zaldarriaga}, {\it {A Hydrodynamical Approach to CMB
  mu-distortions}},  {\em ArXiv e-prints} (2012)
  [\href{http://xxx.lanl.gov/abs/1206.4479}{{\tt arXiv:1206.4479}}].

\bibitem{Weinberg1971}
S.~{Weinberg}, {\it {Entropy Generation and the Survival of Protogalaxies in an
  Expanding Universe}},  {\em \apj} {\bf 168} (Sept., 1971) 175.

\bibitem{wmap9}
{G. Hinshaw et al.}, {\it {Nine-Year Wilkinson Microwave Anisotropy Probe
  (WMAP) Observations: Cosmological Parameter Results}},  {\em ArXiv e-prints}
  (Dec., 2012) [\href{http://xxx.lanl.gov/abs/1212.5226}{{\tt
  arXiv:1212.5226}}].

\bibitem{mabert95}
C.-P. {Ma} and E.~{Bertschinger}, {\it {Cosmological Perturbation Theory in the
  Synchronous and Conformal Newtonian Gauges}},  {\em ApJ} {\bf 455} (Dec.,
  1995) 7--+, [\href{http://xxx.lanl.gov/abs/astro-ph/9506072}{{\tt
  astro-ph/9506072}}].

\bibitem{mukhanov}
V.~F. {Mukhanov} and G.~V. {Chibisov}, {\it {Quantum fluctuations and a
  nonsingular universe}},  {\em Soviet Journal of Experimental and Theoretical
  Physics Letters} {\bf 33} (May, 1981) 532.

\bibitem{kt1995}
A.~{Kosowsky} and M.~S. {Turner}, {\it {CBR anisotropy and the running of the
  scalar spectral index}},  {\em \prd} {\bf 52} (Aug., 1995) 1739,
  [\href{http://xxx.lanl.gov/abs/astro-ph/9504071}{{\tt astro-ph/9504071}}].

\bibitem{dod}
S.~{Dodelson}, {\em {Modern cosmology}}.
\newblock Modern cosmology / Scott Dodelson.~Amsterdam (Netherlands): Academic
  Press.~ISBN 0-12-219141-2, 2003.

\bibitem{ksc2012}
R.~{Khatri}, R.~A. {Sunyaev}, and J.~{Chluba}, {\it {Does Bose-Einstein
  condensation of CMB photons cancel {$\mu$} distortions created by dissipation
  of sound waves in the early Universe?}},  {\em \aap} {\bf 540} (2012) A124.

\bibitem{weinberg}
S.~{Weinberg}, {\em Cosmology}.
\newblock Oxford University Press, Oxford, 2008.

\bibitem{pajer2012}
E.~{Pajer} and M.~{Zaldarriaga}, {\it {New Window on Primordial
  Non-Gaussianity}},  {\em Physical Review Letters} {\bf 109} (2012), no.~2
  021302.

\bibitem{ganc2012}
J.~{Ganc} and E.~{Komatsu}, {\it {Scale-dependent bias of galaxies and
  {$\mu$}-type distortion of the cosmic microwave background spectrum from
  single-field inflation with a modified initial state}},  {\em \prd} {\bf 86}
  (2012), no.~2 023518.

\bibitem{spt2}
{A. van Engelen et al.}, {\it {A Measurement of Gravitational Lensing of the
  Microwave Background Using South Pole Telescope Data}},  {\em \apj} {\bf 756}
  (Sept., 2012) 142, [\href{http://xxx.lanl.gov/abs/1202.0546}{{\tt
  arXiv:1202.0546}}].

\bibitem{act2}
{S. Das et al.}, {\it {Detection of the Power Spectrum of Cosmic Microwave
  Background Lensing by the Atacama Cosmology Telescope}},  {\em Physical
  Review Letters} {\bf 107} (July, 2011) 021301,
  [\href{http://xxx.lanl.gov/abs/1103.2124}{{\tt arXiv:1103.2124}}].

\bibitem{bao1}
{L. Anderson et al.}, {\it {The clustering of galaxies in the SDSS-III Baryon
  Oscillation Spectroscopic Survey: baryon acoustic oscillations in the Data
  Release 9 spectroscopic galaxy sample}},  {\em \mnras} {\bf 427} (Dec., 2012)
  3435--3467, [\href{http://xxx.lanl.gov/abs/1203.6594}{{\tt
  arXiv:1203.6594}}].

\bibitem{bao2}
{C. Blake et al.}, {\it {The WiggleZ Dark Energy Survey: joint measurements of
  the expansion and growth history at z < 1}},  {\em \mnras} {\bf 425} (Sept.,
  2012) 405--414, [\href{http://xxx.lanl.gov/abs/1204.3674}{{\tt
  arXiv:1204.3674}}].

\bibitem{bao3}
{F. Beutler et al.}, {\it {The 6dF Galaxy Survey: baryon acoustic oscillations
  and the local Hubble constant}},  {\em \mnras} {\bf 416} (Oct., 2011)
  3017--3032, [\href{http://xxx.lanl.gov/abs/1106.3366}{{\tt
  arXiv:1106.3366}}].

\bibitem{h01}
{A. Sandage et al.}, {\it {The Hubble Constant: A Summary of the Hubble Space
  Telescope Program for the Luminosity Calibration of Type Ia Supernovae by
  Means of Cepheids}},  {\em \apj} {\bf 653} (Dec., 2006) 843--860,
  [\href{http://xxx.lanl.gov/abs/astro-ph/0603647}{{\tt astro-ph/0603647}}].

\bibitem{h02}
{W. L. Freedman et al.}, {\it {Carnegie Hubble Program: A Mid-infrared
  Calibration of the Hubble Constant}},  {\em \apj} {\bf 758} (Oct., 2012) 24,
  [\href{http://xxx.lanl.gov/abs/1208.3281}{{\tt arXiv:1208.3281}}].

\bibitem{h03}
{A. G. Riess et al.}, {\it {A 3\% Solution: Determination of the Hubble
  Constant with the Hubble Space Telescope and Wide Field Camera 3}},  {\em
  \apj} {\bf 730} (Apr., 2011) 119,
  [\href{http://xxx.lanl.gov/abs/1103.2976}{{\tt arXiv:1103.2976}}].

\bibitem{sn1}
{A. Conley et al.}, {\it {Supernova Constraints and Systematic Uncertainties
  from the First Three Years of the Supernova Legacy Survey}},  {\em \apjs}
  {\bf 192} (Jan., 2011) 1, [\href{http://xxx.lanl.gov/abs/1104.1443}{{\tt
  arXiv:1104.1443}}].

\bibitem{sn2}
{M. Sullivan et al.}, {\it {SNLS3: Constraints on Dark Energy Combining the
  Supernova Legacy Survey Three-year Data with Other Probes}},  {\em \apj} {\bf
  737} (Aug., 2011) 102, [\href{http://xxx.lanl.gov/abs/1104.1444}{{\tt
  arXiv:1104.1444}}].

\bibitem{mm2005}
G.~{Mangano}, G.~{Miele}, S.~{Pastor}, T.~{Pinto}, O.~{Pisanti}, and P.~D.
  {Serpico}, {\it {Relic neutrino decoupling including flavour oscillations}},
  {\em Nuclear Physics B} {\bf 729} (2005) 221--234.

\bibitem{pixie}
A.~{Kogut}, D.~J. {Fixsen}, D.~T. {Chuss}, J.~{Dotson}, E.~{Dwek},
  M.~{Halpern}, G.~F. {Hinshaw}, S.~M. {Meyer}, S.~H. {Moseley}, M.~D.
  {Seiffert}, D.~N. {Spergel}, and E.~J. {Wollack}, {\it {The Primordial
  Inflation Explorer (PIXIE): a nulling polarimeter for cosmic microwave
  background observations}},  {\em \jcap} {\bf 7} (2011) 25.

\bibitem{tegmark}
M.~{Tegmark}, A.~N. {Taylor}, and A.~F. {Heavens}, {\it {Karhunen-Loeve
  Eigenvalue Problems in Cosmology: How Should We Tackle Large Data Sets?}},
  {\em \apj} {\bf 480} (May, 1997) 22,
  [\href{http://xxx.lanl.gov/abs/astro-ph/9603021}{{\tt astro-ph/9603021}}].

\bibitem{matsubara}
T.~{Matsubara}, {\it {Correlation Function in Deep Redshift Space as a
  Cosmological Probe}},  {\em \apj} {\bf 615} (Nov., 2004) 573--585,
  [\href{http://xxx.lanl.gov/abs/astro-ph/0408349}{{\tt astro-ph/0408349}}].

\bibitem{detf}
{A. Albrecht et al.}, {\it {Findings of the Joint Dark Energy Mission Figure of
  Merit Science Working Group}},  {\em ArXiv e-prints} (Jan., 2009)
  [\href{http://xxx.lanl.gov/abs/0901.0721}{{\tt arXiv:0901.0721}}].

\bibitem{sb2002}
R.~{Salvaterra} and C.~{Burigana}, {\it {A joint study of early and late
  spectral distortions of the cosmic microwave background and of the
  millimetric foreground}},  {\em \mnras} {\bf 336} (Oct., 2002) 592--610,
  [\href{http://xxx.lanl.gov/abs/astro-ph/0203294}{{\tt astro-ph/0203294}}].

\bibitem{cosmomc}
A.~{Lewis} and S.~{Bridle}, {\it {Cosmological parameters from CMB and other
  data: A Monte Carlo approach}},  {\em \prd} {\bf 66} (Nov., 2002) 103511,
  [\href{http://xxx.lanl.gov/abs/astro-ph/0205436}{{\tt astro-ph/0205436}}].

\bibitem{wmap}
{D. Larson et al.}, {\it {Seven-year Wilkinson Microwave Anisotropy Probe
  (WMAP) Observations: Power Spectra and WMAP-derived Parameters}},  {\em
  \apjs} {\bf 192} (2011) 16.

\bibitem{spt3}
{C. L. Reichardt et al.}, {\it {A Measurement of Secondary Cosmic Microwave
  Background Anisotropies with Two Years of South Pole Telescope
  Observations}},  {\em \apj} {\bf 755} (Aug., 2012) 70,
  [\href{http://xxx.lanl.gov/abs/1111.0932}{{\tt arXiv:1111.0932}}].

\bibitem{act3}
{S. Das et al.}, {\it {The Atacama Cosmology Telescope: Temperature and
  Gravitational Lensing Power Spectrum Measurements from Three Seasons of
  Data}},  {\em ArXiv e-prints} (Jan., 2013)
  [\href{http://xxx.lanl.gov/abs/1301.1037}{{\tt arXiv:1301.1037}}].

\bibitem{fm2002}
D.~J. {Fixsen} and J.~C. {Mather}, {\it {The Spectral Results of the
  Far-Infrared Absolute Spectrophotometer Instrument on COBE}},  {\em \apj}
  {\bf 581} (2002) 817--822.

\end{thebibliography}\endgroup
\end{document}